\documentclass[aps,prd,showpacs]{revtex4}

\usepackage{amsmath,amscd,amssymb,graphicx}
\usepackage[colorlinks=true,urlcolor=green,linkcolor=black,citecolor=blue]{hyperref}

\begin{document}

\newcommand{\beq}{\begin{equation}}
\newcommand{\eeq}{\end{equation}}

\title{Fermion localization on thick branes}

\author{ Alejandra Melfo, Nelson Pantoja  and Jos\'e David Tempo}

\affiliation{ {\it Centro de F\'{\i}sica Fundamental, Universidad de
Los Andes, M\'erida, Venezuela }}

\begin{abstract}
We consider chiral fermion confinement in scalar thick branes,
which are known to localize gravity, coupled through a Yukawa term.
The conditions for the confinement and their behavior in the
thin-wall limit are found for various different BPS branes,
including double walls and branes interpolating between different
$AdS_5$ spacetimes. We show that only one massless chiral mode is
localized in all these walls, whenever the wall thickness is keep
finite. We also show that, independently of wall's thickness, chiral
fermionic modes cannot be localized in $dS_4$ walls embedded in a
$M_5$ spacetime. Finally, massive fermions in double wall spacetimes
are also investigated. We find that, besides the massless chiral
mode localization, these double walls support quasi-localized
massive modes of both chiralities.  
\end{abstract}


\maketitle

\section{Introduction}

It is a well-known fact that fermions cannot be localized in
Randall-Sundrum \cite{Randall:1999vf} branes. The fermion chiral
modes turn out to be proportional to the inverse warp factor, so
that the same effect that allows to confine gravitons forces the
fermions out of the brane. This can be avoided if the brane is in
fact a domain wall generated by the vacuum expectation value of a
scalar field, as suggested long time ago by Rubakov and Shaposhnikov
\cite{Rubakov:1983bb}, and fermions are coupled to the scalar field.
Bajc and Gabadadze \cite{Bajc:1999mh} showed how this can be
achieved by introducing a Yukawa coupling with the scalar field,
assuming an infinitely thin kink profile for the wall.

However, in order to have a consistent kink profile, the scalar
field should be a solution to the coupled Einstein-scalar field
system with a suitable symmetry breaking potential and the thin
brane must be obtained as the thin wall limit of this solution. Thin
wall geometries have distribution-valued curvatures whose singular
parts are proportional to a Dirac distribution supported on the
surface where the wall is localized, and it is known that strong
conditions must be imposed on a spacetime metric in order to ensure
that its curvature tensors converge (in the sense of distributions)
to the curvature tensors of the limit metric \cite{Geroch:1987qn}.
One of the relevant consequences of these restrictions is that
domain wall solutions cannot be made infinitely thin while keeping
the asymptotic values of the scalar field fixed, as in the
step-function kink of \cite{Bajc:1999mh}.

Localization of gravity on thick domain walls has been considered in
various works
\cite{Gremm:1999pj,DeWolfe:1999cp,Kehagias:2000au,Kakushadze:2000zn,
Wang:2002pk,CBazeia:2003qt,Castillo-Felisola:2004eg} and the thin
wall limit (in the above mentioned distributional convergence sense)
in \cite{Guerrero:2002ki,Melfo:2002wd,Castillo-Felisola:2004eg}.
Some of these thick branes reduce to the Randall-Sundrum (RS) brane
\cite{Randall:1999vf} in the thin-wall limit, but not all of them,
notably dynamic walls \cite{g90,Guerrero:2002ki}, walls that
interpolate between $AdS_5$ spacetimes with different cosmological
constants \cite{Castillo-Felisola:2004eg} and double-walls
\cite{Melfo:2002wd,CBazeia:2003qt}. In some cases, as in the
asymmetric branes found in \cite{gassmukh}, localization of gravity
is not possible \cite{Castillo-Felisola:2004eg}, although the scalar
field behaves as a domain wall and the stress-energy tensor is
distributionally well defined even for thickness zero.

It is therefore of interest to investigate the possibility that
fermion confinement, being directly dependent on the scalar field
solution and not only on the spacetime metric, can be affected by
the internal structure on the thick brane. Confinement of fermions
in general spacetimes has been addressed for example in
\cite{Randjbar-Daemi:2000cr}. Localization of chiral fermions in the
regularized RS scenario of
\cite{DeWolfe:1999cp,Kehagias:2000au,Kakushadze:2000zn} in
\cite{Kehagias:2000au}.  In \cite{Dubovsky:2000am}, it is found that 
 allowing for an 5-dimensional fermion
mass term can result in the confinement of both left and right
modes, although a marginal one.  Namely, fermions can
escape into the bulk by tunneling, and the rate depends on the parameters of the
scalar field potential. Confinement of massive fermions on a numerical thick
brane solution,  has been considered in \cite{Ringeval:2001cq},  introducing an
effective 4-dimensional
mass term on the brane.
 In \cite{Koley:2004at}, the same
results are found using the analytical solution of \cite{Gremm:1999pj}.

In this paper we address the issue of chiral fermion mode
confinement in thick branes, in particular those that exhibit an
interesting internal structure. After reviewing this confinement in
two BPS regularized versions of the RS brane
\cite{Kehagias:2000au,Koley:2004at}, we consider the asymmetric BPS
branes \cite{Castillo-Felisola:2004eg}, where the wall separates
spacetimes with different cosmological constants, and find the
conditions for fermion mode localization. The Yukawa coupling
required to localize chiral fermions is shown to diverge in the
thin-wall limit for all these BPS branes. We then turn to spacetimes
with a $dS_4$ thick brane embedded in a $M_5$ bulk, whose four
dimensional version was found in \cite{g90}. These are known to
localize gravity \cite{Wang:2002pk,Kehagias:2002qk} and could be of
cosmological interest. However, we find that these walls cannot
localize fermion modes, independently of the wall's thickness.
Finally, we explore the double walls found in \cite{Melfo:2002wd}.
We show that massless chiral fermions can be confined in these
walls, and that in the massive case they can allow for left- and
right-handed mode quasi-confinement.

\section{Localization of massless fermions}

Let us first consider a domain wall solution of the coupled
Einstein-scalar fields equations, with an asymptotically $AdS_5$
metric of the form \beq \label{propermetric} ds^2 = e^{2 A(r)
}\eta_{\mu\nu} dx^\mu dx^\nu + dr^2  \eeq where $\eta$ is the
4-dimensional Minkowski metric (in the following a capital index
like $M$ runs from 0 to 4 while greek indices run from 0 to 3, we
use units where $8\pi G_5 =1$, with $G_5$ the gravitational constant
in five dimensional spacetime).

The 5-dimensional spinor field satisfies the massless Dirac equation
in the background (\ref{propermetric}) \beq \Gamma^M \nabla_M
\Psi(x,r) = 0 \eeq With the decomposition \beq \Psi(x,r)_{\pm} =
\psi(x)_{\pm} f(r)_\pm \eeq and demanding $\psi(x)_{\pm}$ satisfy
the massless 4-dimensional Dirac equation, the equation for the
chiral modes is integrated to give \cite{Bajc:1999mh} \beq f(r)_\pm
\propto e^{- 2 A(r)} \eeq So that the very fact that the
gravitational modes are confined to the wall, i.e. $e^{2 A(r) } \to
0$ as $|r| \to \infty$, implies that $f(r)_\pm$ is not normalizable,
i.e. that the fermionic fields cannot be confined. In order to have
fermions on the wall, one can couple them to the scalar field with a
Yukawa term \cite{Bajc:1999mh} of the form $\lambda \overline{\Psi}
\Psi \phi$, and then get \beq f(r)_\pm \propto e^{- 2 A(r) \pm
\lambda \int{\phi(r) dr}} \label{fmodes} \eeq For appropriate values
of $\lambda  $, the function $f(r)_-$ can in principle be normalized
and one chiral fermion mode confined. Let us consider some specific
solutions for the domain wall $\phi(r)$

\subsection{The Kink} The solution \beq\phi
= \phi_0\tanh \left(\frac{\alpha r}{\delta}\right) ; \qquad \phi_0
\equiv \sqrt{{3\delta}} \label{kink} \eeq for the Einstein-scalar
field system with a symmetry breaking potential $V(\phi)$ is found
when \cite{DeWolfe:1999cp} \beq \label{kinkA} A(r) = -
\frac{2}{3}\delta \left[ \ln\left(\cosh\left(\frac{\alpha r}{\delta}
\right) \right) + \frac{1}{4} \tanh^2\left(\frac{\alpha r}{\delta}
\right)\right]
 \eeq with \beq\label{kinkV} V(\phi) =
\frac{9}{2}\alpha^2\left[\frac{1}{\phi_0^2}\left(1 -
\frac{\phi^2}{\phi_0^2}\right)^2 - \frac{4}{3} \left(
\frac{\phi}{\phi_0}\right)^2\left(1 -\frac{1}{3}
\left(\frac{\phi}{\phi_0}\right)^2\right)^2\right] \eeq and the
$AdS_5$ cosmological constant is given by $\Lambda = - 8
\alpha^2/3$. Notice that although $V(\phi)$ as given by
(\ref{kinkV}) is unbounded from below, a stability argument exists
for this kind of BPS walls \cite{Skenderis:1999mm}. Confinement of
gravitational modes for this model, in different parameterized
forms, is discussed in \cite{Kehagias:2000au,Kakushadze:2000zn}.

The thin-wall limit is found for $\delta \to 0$ and corresponds to
$A(r) \to -2\alpha |r|/3$, which behaves as the usual warp factor of
the RS geometry, while $\phi\rightarrow 0$ in this limit. Actually,
following \cite{Guerrero:2002ki}, it can be proved that
(\ref{propermetric},\ref{kinkA}) provides a sequence of metrics that
satisfies the required convergence conditions of
\cite{Geroch:1987qn}. Then the distributional limit $\delta \to 0$
of all the curvature tensor fields of
(\ref{propermetric},\ref{kinkA}) exists and gives the curvatures of
the limit metric. This rather technical proof will not be presented
here.

Now, for this domain wall, it is readily found using (\ref{fmodes})
that one chiral fermion mode is localized on the brane when \beq
\lambda > \sqrt{\frac{|\Lambda|}{2\delta}} \eeq This result, in a
slightly different parameterized form, has been found in
\cite{Kehagias:2000au}. It should be noted that, strictly speaking,
there is no confinement in the infinitely-thin wall.

\subsection{Domain wall for a sine-Gordon
potential}\label{sine-wall}
A solution with \beq \phi = \sqrt{3 \delta} \arctan
\left(\sinh\left( \frac{\alpha r}{\delta}\right) \right) \label{RS}
\eeq can be found \cite{Gremm:1999pj} for the metric
(\ref{propermetric}) with \beq A(r) = -\delta\ln \left(\cosh\left(
\frac{\alpha r}{\delta}\right) \right) \label{RSA} \eeq and the
scalar field potential \beq V = 3 \alpha^2 \left[\left(\frac{1}{2
\delta}+ 2 \right) \cos^2\left(\frac{\phi}{\sqrt{3\delta}}\right) -
2\right] \eeq The parameter $\alpha$ is given by the cosmological
constant, $\Lambda = - 6 \alpha^2$. This smooth domain wall geometry
(parameterized in a sligtly different form) has been shown to
localize four-dimensional gravity on the wall \cite{Gremm:1999pj}.

The thin-wall limit for this solution is found for $\delta \to 0$
and corresponds to $A(r) \to -\alpha |r|$, the usual warp factor of
the RS geometry. It can be shown that in this limit the
energy-momentum tensor and all the curvatures converge rigourously
to the corresponding tensor distributions associated to the RS thin
brane geometry \cite{Guerrero:2002ki,Pantoja:2003zr}), with singular
parts which are proportional to a $\delta$-distribution supported on
the wall's plane. Now, it should be noticed that the scalar field
(\ref{RS}) vanishes everywhere in the thin-wall limit $\delta \to
0$. Since $\delta$ is the only parameter that can be consistently
interpreted as the wall's thickness, it is not possible to take the
thin-wall limit rigorously while keeping $\phi\neq 0$.

Now, equation (\ref{fmodes}) gives \beq f(r)_{\pm} \propto
\left(\cosh\left( \frac{\alpha r}{\delta}\right) \right)^{2 \delta}
 \exp\left\{
\pm \lambda \sqrt{3}\,\frac{\delta^{3/2}}{\alpha}\sum_{k=0}^{\infty}
\frac{E_k \left[\arctan (\sinh \alpha r/{\delta})
\right]^{2k+2}}{(2k+2)(2k)!}\right\} \eeq where $E_k$ are the
Euler's numbers ($E_1=1$, $E_2=5$, $E_3=61$, etc. and $E_0=1$),
which can be rewritten as

\beq \label{fmodes2}f(r)_{\pm} \propto
\exp\left\{\sum_{k=1}^{\infty}\left[ \frac{\delta
\,2^{2k}\,(2^k-1)\,B_k}{k\,(2k)!} \pm \frac{\lambda\,
\sqrt{3}\,\delta^{3/2}\,E_{k-1} }{2\,\alpha\,
k\,(2k-2)!}\right]\left[\arctan (\sinh \alpha r/{\delta})
\right]^{2k}\right\} \eeq where $B_k$ are the Bernoulli's numbers
($B_1=1/6$, $B_2=1/30$, $B_3=1/42$, etc.) From (\ref{fmodes2}) it
follows that in order to confine one of the chiral modes, one must
require \beq\label{exactly} \lambda > \frac{2\alpha}{3\,\sqrt{ 3
\delta}}= \frac{1}{3^2}\sqrt{\frac{2|\Lambda|}{\,\delta}} \eeq
Clearly in this model, as in the previous one, there is no
confinement in the infinitely-thin wall.

Localization of fermions on this thick brane, with a different
parametrization and employing the asymptotic values of $\phi$ as
$r\rightarrow \pm \infty$ \cite{Bajc:1999mh} in (\ref{fmodes})
instead of evaluating explicitly the integral as we have done, is
discussed in \cite{Koley:2004at}. Here it is instructive to compare
the result obtained in this way with the exact result
(\ref{exactly}). In the large $r$ limit, we find that asymptotically
\beq f(r)_{\pm} \propto \exp \left( 2 \alpha  \pm \lambda\,\sqrt{ 3
\delta}\,\frac{\pi}{2}\right) |r| \eeq  so that in order to confine
one of the chiral modes, one must have \beq \lambda >
\frac{4\alpha}{\pi \sqrt{ 3 \delta}} = \frac{2}{3\pi}\sqrt{\frac{2
|\Lambda|}{\delta}} \eeq

Notice that solutions to the Einstein-scalar field equations with a
plane-parallel symmetry with a given $V(\phi)$ and $\phi(r)$ are
always part of a family of ``shifted'' solutions with $\tilde
\phi(r) = \phi(r) - \epsilon $ and $\tilde V(\tilde \phi) = V(\tilde
\phi + \epsilon ) $, with $\epsilon $ a constant, having the same
metric and cosmological constant. The shifting is irrelevant for the
gravitational properties, but not for the fermion confinement, as is
easily seen by setting $\epsilon \to \phi(r=\infty)$: fermion modes
can then escape to infinity, or in other words, the Yukawa coupling
required is infinite. In the case of wall separating  spacetimes
with the same cosmological constant, the Yukawa coupling required is
minimal for $\epsilon=0$, the symmetric kink solution. But this is
not the case for asymmetric walls,  as we consider now.

\subsection{Asymmetric walls}
An example of an asymmetric domain wall was found in
\cite{Castillo-Felisola:2004eg}, with \beq \phi =
2\sqrt{3\delta}\left[ \exp\{ - e^{-\beta r/\delta}\} -
\epsilon\right] \eeq \beq V(\phi)= 18\left\{ \left[ \frac{\beta}{12
\delta} {\tilde\phi}\ln\left(\frac{{\tilde\phi}^2}{12 \delta}\right)
\right]^2 -  \left[\frac{\beta}{12 \delta}{\tilde\phi}^2 \left(1 -
\ln\left(\frac{{\tilde\phi}^2}{12 \delta}\right) \right) - \alpha
\right]^2\right\},  \qquad \tilde\phi =  \phi + 2\sqrt{3\delta}
\epsilon \label{albapot} \eeq and a warp factor \beq A(r)= \alpha r
- \delta \exp(-2 e^{-\beta r /\delta } ) + \delta\, {\rm
Ei}\left(-2e^{-\beta r/\delta}\right)  , \label{albamu} \eeq where
the exponential integral is given by $ {\rm Ei}(u) \equiv
-\int_{-u}^\infty e^{-\tau}/\tau\, d\tau $ (see
\cite{Castillo-Felisola:2004eg} for details). This is a  family of
domain wall spacetimes without reflection symmetry along the
direction perpendicular to the wall. Far from the wall, the warp
factor is \beq A(r\to -\infty) = \alpha \, r , \qquad A(r\to \infty)
=  - (\beta -\alpha) \, r \eeq The wall thus interpolates between
$AdS_5$ spacetimes with cosmological constant $\Lambda_- = - 6
\alpha^2$ for $r < 0$ and $\Lambda_+ = -6 (\beta - \alpha)^2 $ for
$r > 0$. The constant $\epsilon$ fixes the asymptotic values of the
scalar field at $r = \pm \infty$ and $\delta$ is the wall's
thickness. This asymmetric domain wall geometry (for
$\beta>\alpha>0$) has been shown to localize four-dimensional
gravity on the wall \cite{Castillo-Felisola:2004eg}.

Now, equation (\ref{fmodes}) gives \beq f(r)_{\pm} \propto
-2\left[\alpha r - \delta \exp(-2 e^{-\beta r /\delta } ) + \delta\,
{\rm Ei}\left(-2e^{-\beta r/\delta}\right) \right]\pm
2\,\lambda\,\sqrt{3\delta}\left[-\frac{\delta}{\beta}\,{\rm
Ei}\left(-e^{-\beta r/\delta}\right)-\epsilon\,r \right]\eeq whose
behavior is however rather complicated. Again, one can study the
asymptotic behavior of the chiral modes. We have for the $\phi$
field \beq \phi(r\to -\infty) = - 2 \sqrt{3\delta} \epsilon \qquad
\phi(r\to\infty) = - 2 \sqrt{3\delta}(1- \epsilon)\eeq The Yukawa
coupling required to confine fermions to these walls is minimized
when \beq \epsilon^{-1} =  1 + \sqrt{|\Lambda_+| / |\Lambda_-|} \eeq
and the condition for this confinement reads \beq \lambda >
\frac{1}{3\sqrt{2\delta}} (\sqrt{|\Lambda_+|} + \sqrt{|\Lambda_-|} )
\eeq Since the scalar field depends on  $\beta \propto
\sqrt{|\Lambda_+|} + \sqrt{|\Lambda_-|}$, fermions are confined to
the wall as long as one of the cosmological constants is non-zero;
of course gravity is not localized if $\Lambda_+=0$ or
$\Lambda_-=0$.

\subsection{A de Sitter thick brane}
A domain wall solution interpolating between spacetimes without a
5-dimensional cosmological constant can be found if one allows for a
de Sitter expansion on the wall plane, i.e. a dynamic solution
\cite{g90}, that can be of cosmological interest. The
Einstein-scalar field system can be integrated for a metric \beq
ds^2 = e^{2 A(\bar r)}(- dt^2 + e^{2 \beta t}\,dx^i\,dx^i +
d\bar{r}^2  ) \label{conformetric} \eeq where $i=1,\,2,\,3$.

The scalar field is given by \beq \phi(\bar{r}) = \sqrt{3 \delta (1
- \delta)} \arctan \left( \sinh\left(\frac{\beta \bar{r}}{\delta}
\right)\right) \label{phig90} \eeq with \beq A(\bar r) = -\delta \ln
\left( \cosh\left(\frac{\beta \bar r}{\delta} \right)\right) \eeq
and a sine-Gordon scalar field potential. These walls have a
well-defined distributional thin limit $\delta \rightarrow 0$, a
thin $dS_4$ brane embedded in a $M_5$ bulk
\cite{Guerrero:2002ki,Pantoja:2003zr}). Although they do not become
an RS brane in the thin wall limit, they can localize gravity with a
mass gap between the zero modes and the massive ones
\cite{Wang:2002pk,Kehagias:2002qk,Castillo-Felisola:2004eg}.
However, they cannot confine fermions as we now show.

In these coordinates, the Dirac equation results in a different
expression for $f_\pm$ \beq \label{confmodes}f_\pm \propto
\exp\left\{- 2 A(\bar r) \pm \lambda \int{e^{A(\bar r)} \phi(\bar r)
d\bar r }\right\} \eeq which can be explicitly integrated and yields
a power series of $\arctan (\sinh \alpha r/{\delta})$ as in
(\ref{fmodes2}), but quite more involved with coefficients which are
polynomials in $\delta$. Each term of the series, for increasing
power, pushes up the lower bound on $\lambda$ necessary to obtain
the desired normalizability of $f_{\pm}$. Hence, there is no value
of $\lambda$ that renders $f_\pm$ normalizable. This is not at all
unexpected, in this model $A(\bar r)$ is the same function $A(r)$ of
the domain wall with a sine-Gordon potential and $f_\pm$ can be
rewritten as \beq \label{ypsilon}f_\pm \propto \exp{\int{d\bar r}\,
\Upsilon(\bar r)_{\pm}} \eeq where \beq \Upsilon(\bar r)_{\pm}=
\left[ 2\,\alpha\,\tanh \frac{\alpha \bar r}{\delta} \pm \lambda
\,\sqrt{3\delta(1-\delta)} \,\exp({-\delta\ln (\cosh\frac{\alpha
\bar r}{\delta})})\, \arctan(\sinh \frac{\alpha \bar
r}{\delta})\right] \eeq It follows that the Yukawa coupling
contributes to $\Upsilon(\bar r)_{\pm}$ with a term that becomes
exponentially suppressed with respect to the one of the domain wall
of the sine-Gordon potential, becoming unable to cancel the coming
one from $\exp \{-2\, A(\bar r)\}$ for $|\bar r| \to \infty$.
Asymptotically, \beq \Upsilon(\bar r)_{-} \to \pm \left(2 \alpha -
\lambda \sqrt{3 \delta (1 - \delta)}\,e^{-\beta |\bar r}|
\,\frac{\pi}{2}\right) \eeq as $\bar r\to \pm \infty$ and the term
coming from the Yukawa coupling goes to zero. This behavior has also
been checked by numerical integration of (\ref{confmodes}).

\subsection{Double walls}
In \cite{Melfo:2002wd}, a two-parameter family of walls are found
with a metric \beq ds^2 = e^{2 A(\bar r) } \left( \eta_{\mu \nu}
dx^\mu dx^\nu + d\bar r^2 \right ) \label{skirzmetric} \eeq The
scalar field solution is \beq \phi(\bar r) =\phi_0 \arctan(\alpha
\bar r)^{s} \, ; \; \phi_0 = \frac{\sqrt{6 s - 3}}{s} \label{skirz}
\eeq with \beq A(\bar r) = -\frac{1}{2s} \ln\left( 1 + (\alpha \bar
r)^{2s}\right) \label{skirzA} \eeq and \beq V = 3 \alpha^2
\sin(\phi/\phi_0)^{2-2/s}\left[ \frac{2 s + 3}{2}\cos^2(\phi/\phi_0)
- 2 \right] \eeq If $s$ is an odd integer, the field (\ref{skirz})
interpolates between two global minima of the potential, but for $s
> 1$ there is a local minima between any two global ones. The
configuration then is that of a double wall, with the energy density
peaking at two points, interpolating between $AdS_5$ asymptotic
vacua with $\Lambda = - 6 \alpha^2$. These walls have been shown to
confine gravity in \cite{Castillo-Felisola:2004eg}; similar double
walls have been found in \cite{CBazeia:2003qt}. For $s=1$, this is
just the RS regularized spacetime considered in \ref{sine-wall}, for
a fixed thickness $\delta=1$ and written in conformal coordinates.

The chiral modes take again the form (\ref{confmodes}), but in
contrast with the dynamic walls, confinement of chiral fermions is
possible in this case, for any odd value of $s$. The integration of
(\ref{confmodes}) in this case has again been performed numerically,
giving a renormalizable mode. But it is also possible in this case
to examine the asymptotic behavior of the integrand, writing the
modes as in (\ref{ypsilon}) we find \beq \Upsilon(\bar r)_{\pm} \to
\left(\frac{2}{\bar r} \pm \lambda \frac{\pi}{2}\frac{\sqrt{6s
-3}}{s} \frac{1}{\alpha \bar r}\right) \eeq as $|\bar r|\to \infty$.
So that the Yukawa term dominates when \beq \lambda >
\frac{4\alpha}{\pi\sqrt{3}} \frac{s}{\sqrt{2s - 1}} \eeq

\section{Localization of massive fermions}

Up to now, we have supposed that the 4-dimensional spinor fields
satisfy the massless Dirac equation on the wall. The inclusion of a
mass term that connects left and right-handed modes could  modify
the well-known result of having only one chiral mode localized on
the brane. This possibility has been explored in Ref.
\cite{Ringeval:2001cq}, using a numerical solution for the domain
wall.  If instead of this one uses the exact solutions
(\ref{RS},\ref{RSA})  as in Ref. \cite{Koley:2004at}, or
(\ref{kink},\ref{kinkA}), very similar results are found. Namely,
the chiral modes can be shown to satisfy a Schr\"odinger-like
equation, with different potentials for left and right-modes. These
potentials are unbounded from below, but have minima where the modes
can be confined, and the tunneling can in  principle be sufficiently
suppressed by adjusting the parameters.

Here we study  the case of the  double wall spacetime of Ref.
\cite{Melfo:2002wd}, which includes the regularized version of the
RS geometry considered in \cite{Gremm:1999pj} as a particular case.
These branes have the advantage of being written in a coordinate
system that leads to a Schr\"odinger equation for the chiral modes
with the four-dimensional fermion masses as its eigenvalues, which
enable us to analyze its solutions in a simpler way.

\begin{figure}
\includegraphics[height=6cm]{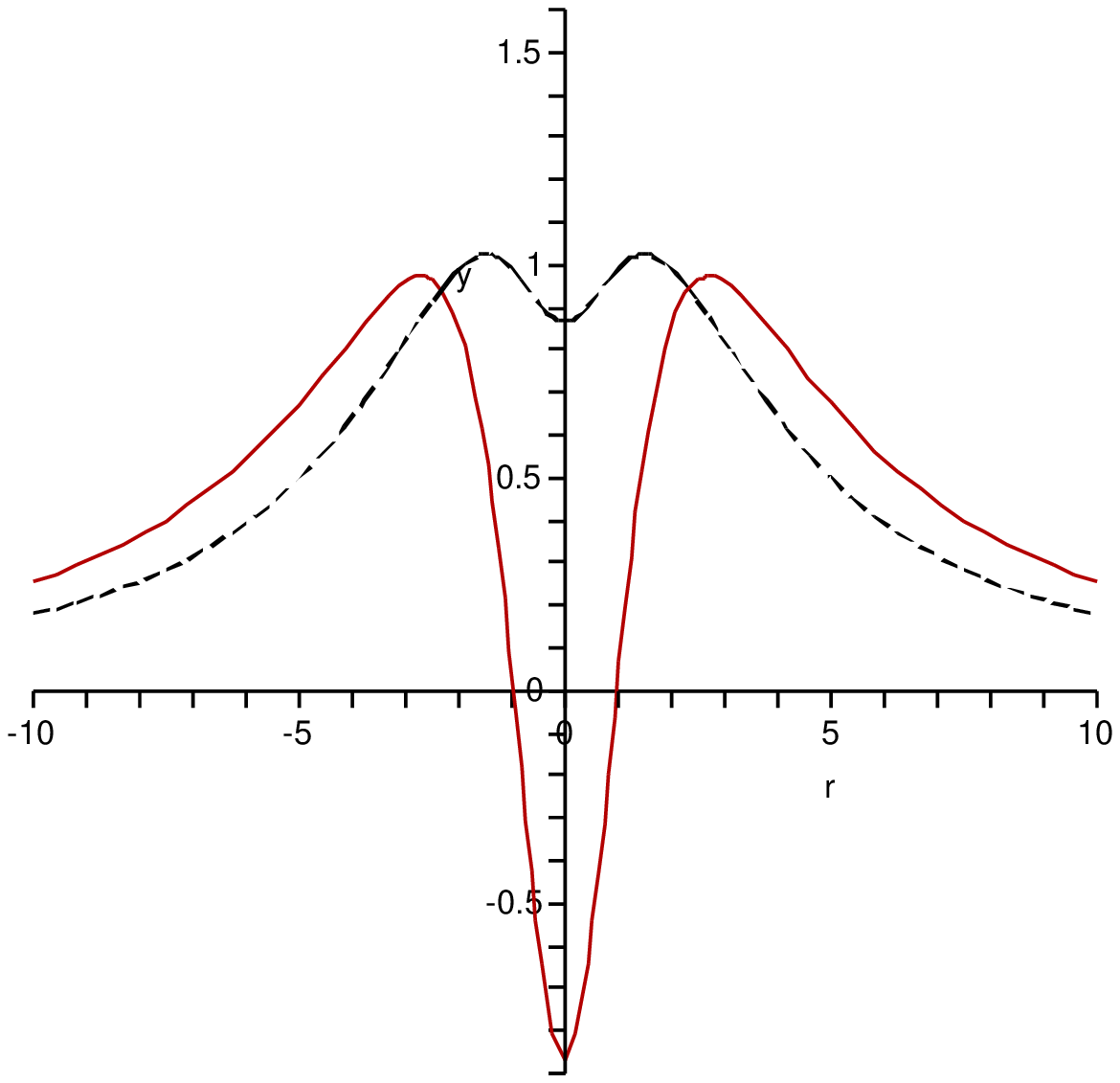}\includegraphics[height=6cm]{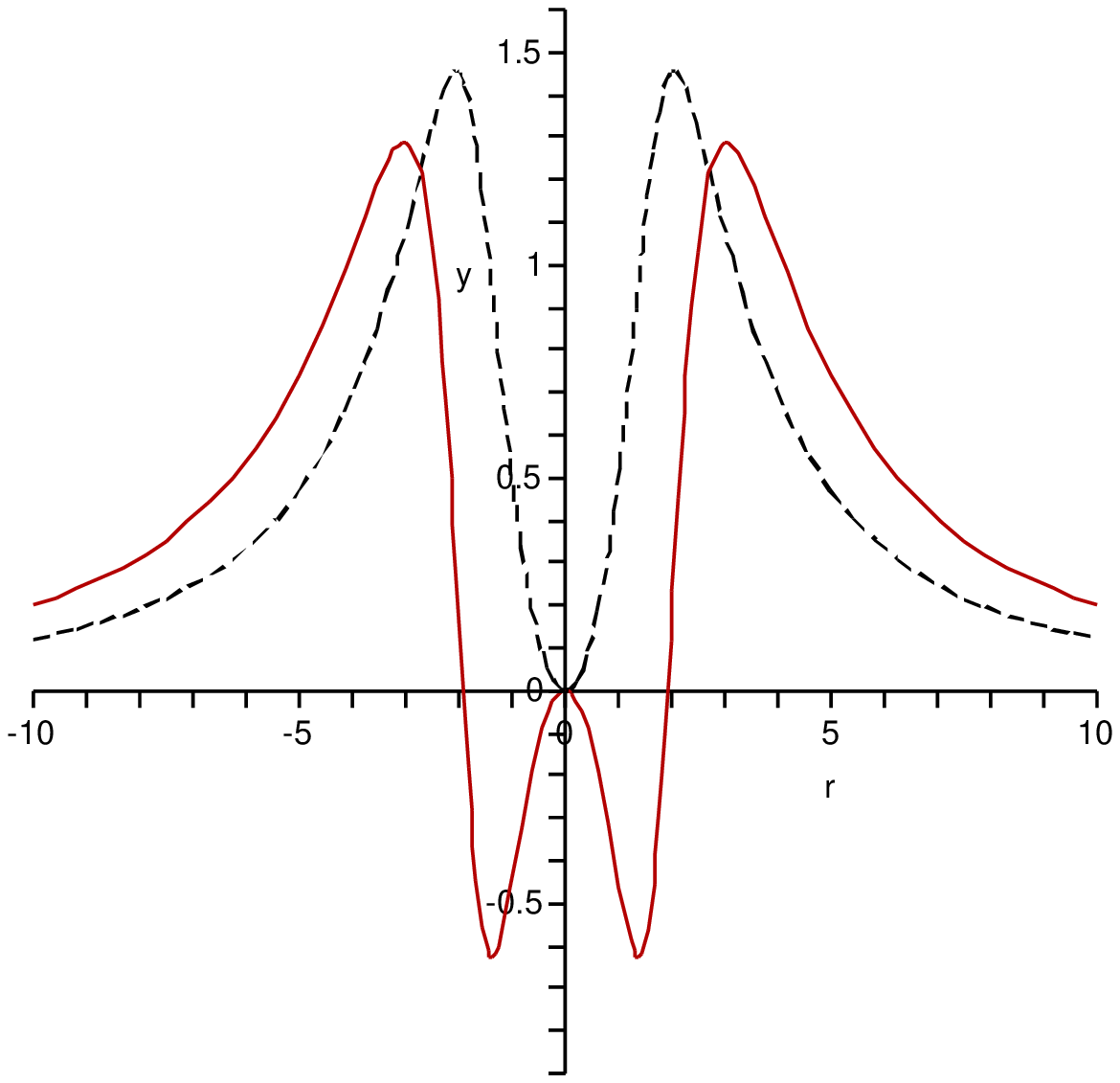}\includegraphics[height=6cm]{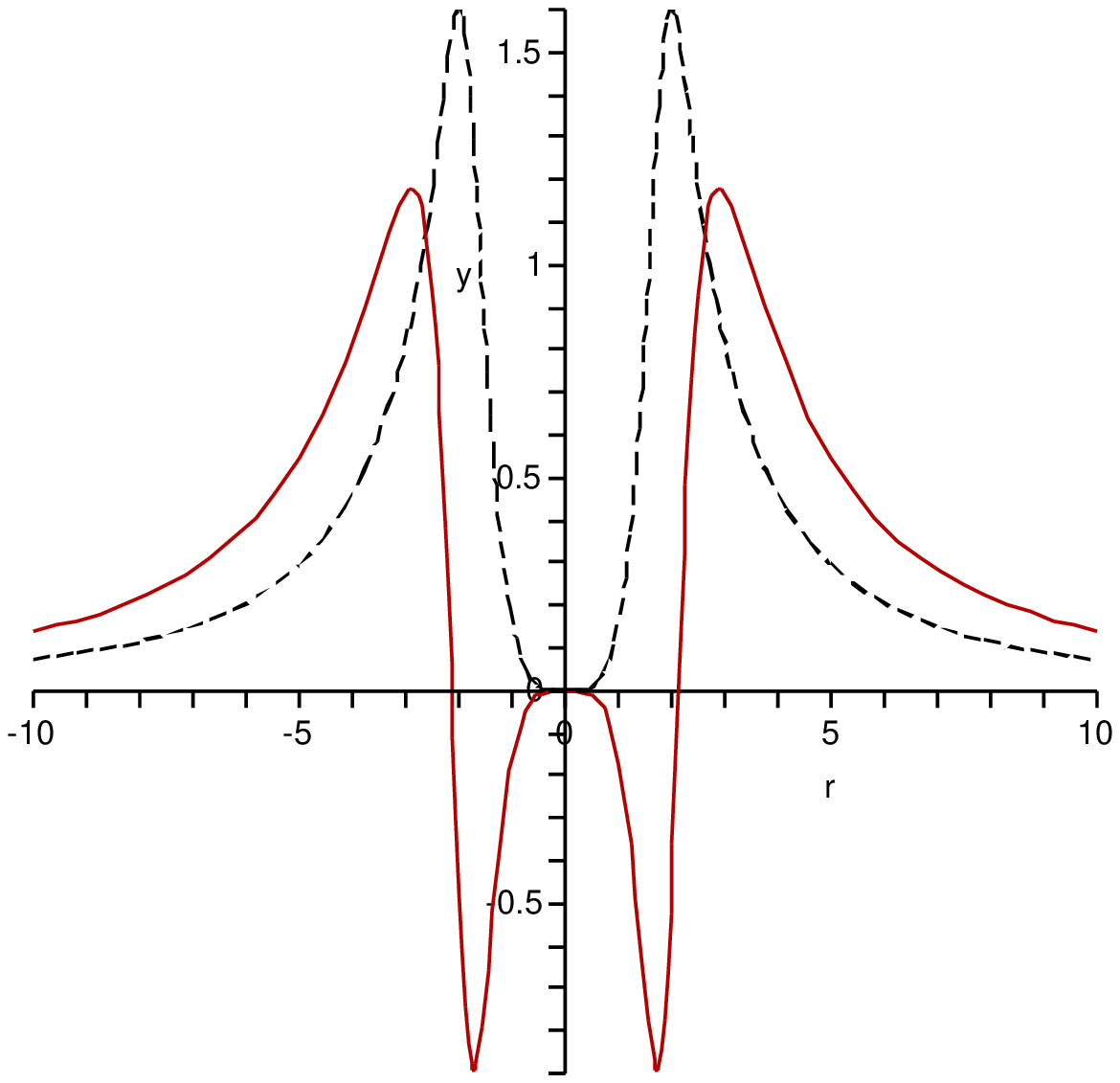}
\caption{ $V_{QM}^{-}$ (continuous line) and  $V_{QM}^{+}$ (dashed
line) for the double walls with $s=1$ , $s=3$ and $s=5$ }
\label{fig2}
\end{figure}

Consider again a spinor $\Psi$ in 5 dimensions, \beq \Psi(x,\bar r)
= \psi_+(x) u_+(\bar r) + \psi_-(x) u_-(\bar r) \eeq where now  the
4-dimensional spinors satisfy \beq \label{4-Dirac} i \gamma^\mu
\nabla_\mu \psi_\pm = m \psi_\mp \eeq Then the chiral modes follow
the equations (prime denotes derivative respect to $\bar r$) \beq
\left[
\partial_r + 2 A'(\bar r) \pm \lambda \phi(\bar r)
 e^{A(\bar r)} \right] u_{\mp} = \pm \,m \,u_{\pm}
\eeq Setting \beq \label{u-mass} u(\bar r) = \hat u(\bar r) e^{- 2
A(\bar r)} \eeq one can write the Schr\"odinger equations \beq
\label{S-E} \left[ -\partial^2_r + V_{QM}^{\pm}   \right]
\hat{u}_{\pm} =  m^2 \hat{u}_{\pm} \eeq where \beq \label{VQM}
V_{QM}^{\pm} =  (\lambda \phi(\bar r) e^{A(\bar r)} )^2 \pm
\partial_r (\lambda \phi(\bar r) e^{A(\bar r)}) \eeq In figure
\ref{fig2} we plot $V_{QM}^{\pm}$ for different values of $s$. In
all cases $\alpha =1$, $\lambda = 1$.

Now, notice that (\ref{S-E}) can be written as \beq \label{SUSY-QM}
Q^+ Q\,\hat{u}_-= m^2\,\hat{u}_-,\qquad Q\, Q^+\,\hat{u}_+=
m^2\,\hat{u}_+\eeq where $Q\equiv\partial_{\bar r} +
\lambda\,\phi(\bar r)\, \exp A(\bar r)$ and $Q^+\equiv
-\partial_{\bar r} + \lambda\,\phi(\bar r)\, \exp A(\bar r)$, and
(\ref{S-E}) can be recast as a SUSY quantum mechanics problem as in 
\cite{Arkani-Hamed:1999dc}. It follows that the eigenvalues of the
$\hat{u}_-$ and $\hat{u}_+$ modes always come in pairs, except
possibly for the massless modes. Indeed, this pairing of the mass
eigenmodes is required for the existence of massive fermions
satisfying (\ref{4-Dirac}).

As a simple check of consistency, let us consider the massless modes
again. For $m=0$, from (\ref{SUSY-QM}) we find \beq
\hat{u}_{0\pm}\propto \exp \{ \pm\lambda\int d\bar
r\,\phi\,e^A\}\eeq  and from (\ref{u-mass}) the result
(\ref{confmodes}) is recovered. Since for \beq \lambda >
\frac{4\alpha}{\pi\sqrt{3}} \frac{s}{\sqrt{2s - 1}} \eeq one and
only one chiral mode is normalizable, this chiral normalizable mode
is just the one that was considered in the previous section and that
gives a chiral four dimensional fermion localized on the brane.

Now let us consider the massive modes. We recall that for $s=1$,
this is just the second RS regularized spacetime considered in the
previous section, for a fixed thickness $\delta=1$ and written in
conformal coordinates. For $s=1$, we see from Fig. \ref{fig2} that
  right- and left-handed  modes are  trapped in a
metastable state with a non-zero probability of tunneling into the
bulk. This tunneling probability depends on the value of the Yukawa
coupling and could be made small by increasing $\lambda$.  For the range of 
parameters
shown, this tunneling probability is very high, to the point that one can exclude localization. 
However, for the same set of parameters but  greater values of $s$, where double walls 
appear, we
see from Fig. \ref{fig2} that $V_{QM}^{+}$ develops a deep minimum in the
region between the walls. Now, right-handed modes of small mass can
be trapped here in a metastable state like the paired massive
left-handed ones. As before, the tunneling probability can be made
small by increasing $\lambda$. On the other hand, increasing $s$ has
the effect of making each of the walls thinner, accordingly,
$V_{QM}^{+}$ has also thinner ``walls'', which would increase the
probability of fermions escaping. A precise calculation of the
tunneling probability for these massive left and right-handed modes
is in order here, but will only make sense in a definite model in
which the parameters of the potential can be related with the age of
the Universe. This is beyond the scope of this paper and will be
dealt with elsewhere \cite{grmp}. It is however clear that the internal
 structure of the wall,
i.e. the appeareance of double walls, can enhance the probability of localizing
 massive fermions.

\section{Summary}
We have studied the localization of fermions on thick branes with a
non-trivial internal structure. We find that double walls can
confine fermions of a given chirality, as is also the case with the
known regularized RS branes, provided they are coupled to the scalar
field through a Yukawa coupling that is inversely proportional to
the wall's thickness. Asymmetric BPS walls interpolating between
spacetimes with different cosmological constants can  confine
fermions in similar way, with a Yukawa coupling proportional to the
sum of the square roots of the cosmological constants.  In all these
cases, the Yukawa coupling required diverges in the thin-wall limit.
On the other hand, dynamic walls with a cosmological de Sitter
expansion are unable to localize even one chiral mode. We have
considered here only the five dimensional version of the well-known
solution of Goetz \cite{g90}, but these results are expected to hold
for other dynamic walls such as the ones found recently in
\cite{Guerrero:2005aw}; work in this direction is now in progress
\cite{grmp}.

Finally, we have also addressed the issue of confinement of massive
fermions in double walls.  We find that,
besides the massless chiral mode localization, massive modes of both
chiralities can be quasi-localized on double walls, being ``trapped''
in between the walls. The tunneling probability can in principle be
made as small as desired by increasing the Yukawa coupling and
fermion masses, but for a given set of parameters it is smaller in the double
 walls than in the
single ones. Thus, fermion localization is shown to be affected by the internal
 structure of the
domain wall solutions.

\section{Acknowledgments}

We wish to thank B. Bajc, R. Guerrero and R. O. Rodr\'iguez for discussions.
This work was partially financed by CDCHT-ULA project No.
C-1267-04-05-A and  by FONACIT projects S1-2000000820 and
F-2002000426.


\begin{thebibliography}{999}



\bibitem{Randall:1999vf}
  L.~Randall and R.~Sundrum,
  %``An alternative to compactification,''
  Phys.\ Rev.\ Lett.\  {\bf 83} (1999) 4690
  [arXiv:hep-th/9906064].
  %%CITATION = HEP-TH 9906064;%%
  See also
  M.~Gogberashvili,
  %``Hierarchy problem in the shell-universe model,''
  Int.\ J.\ Mod.\ Phys.\ D {\bf 11} (2002) 1635
  [arXiv:hep-ph/9812296].
  %%CITATION = HEP-PH 9812296;%%

%\cite{Rubakov:1983bb}
\bibitem{Rubakov:1983bb}
  V.~A.~Rubakov and M.~E.~Shaposhnikov,
  %``Do We Live Inside A Domain Wall?,''
  Phys.\ Lett.\ B {\bf 125} (1983) 136.
  %%CITATION = PHLTA,B125,136;%%

 %\cite{Bajc:1999mh}
\bibitem{Bajc:1999mh}
  B.~Bajc and G.~Gabadadze,
  %``Localization of matter and cosmological constant on a brane in anti de
  %Sitter space,''
  Phys.\ Lett.\ B {\bf 474} (2000) 282
  [arXiv:hep-th/9912232].
  %%CITATION = HEP-TH 9912232;%%

%\cite{Geroch:1987qn}
\bibitem{Geroch:1987qn}
  R.~Geroch and J.~H.~Traschen,
  %``Strings And Other Distributional Sources In General Relativity,''
  Phys.\ Rev.\ D {\bf 36} (1987) 1017.
  %%CITATION = PHRVA,D36,1017;%%

%\cite{Gremm:1999pj}
\bibitem{Gremm:1999pj}
  M.~Gremm,
  %``Four-dimensional gravity on a thick domain wall,''
  Phys.\ Lett.\ B {\bf 478} (2000) 434
  [arXiv:hep-th/9912060].
  %%CITATION = HEP-TH 9912060;%%

%\cite{DeWolfe:1999cp}
\bibitem{DeWolfe:1999cp}
  O.~DeWolfe, D.~Z.~Freedman, S.~S.~Gubser and A.~Karch,
  %``Modeling the fifth dimension with scalars and gravity,''
  Phys.\ Rev.\ D {\bf 62} (2000) 046008
  [arXiv:hep-th/9909134].
  %%CITATION = HEP-TH 9909134;%%

%\cite{Kehagias:2000au}
\bibitem{Kehagias:2000au}
  A.~Kehagias and K.~Tamvakis,
  %``Localized gravitons, gauge bosons and chiral fermions in smooth spaces
  %generated by a bounce,''
  Phys.\ Lett.\ B {\bf 504} (2001) 38
  [arXiv:hep-th/0010112].
  %%CITATION = HEP-TH 0010112;%%


%\cite{Kakushadze:2000zn}
\bibitem{Kakushadze:2000zn}
  Z.~Kakushadze and P.~Langfelder,
  %``Gravitational Higgs mechanism,''
  Mod.\ Phys.\ Lett.\ A {\bf 15} (2000) 2265
  [arXiv:hep-th/0011245].
  %%CITATION = HEP-TH 0011245;%%

%\cite{Wang:2002pk}
\bibitem{Wang:2002pk}
  A.~z.~Wang,
  %``Thick de Sitter brane worlds, dynamic black holes and localization of
  %gravity,''
  Phys.\ Rev.\ D {\bf 66} (2002) 024024
  [arXiv:hep-th/0201051].
  %%CITATION = HEP-TH 0201051;%%

\bibitem{CBazeia:2003qt}
  D.~Bazeia, C.~Furtado and A.~R.~Gomes,
  %``Brane structure from scalar field in warped spacetime,''
  JCAP {\bf 0402} (2004) 002
  [arXiv:hep-th/0308034].
  %%CITATION = HEP-TH 0308034;%%
  See also D.~Bazeia, J.~Menezes and R.~Menezes,
  %``New global defect structures,''
  Phys.\ Rev.\ Lett.\  {\bf 91} (2003) 241601
  [arXiv:hep-th/0305234].
  %%CITATION = HEP-TH 0305234;%%

%\cite{Castillo-Felisola:2004eg}
\bibitem{Castillo-Felisola:2004eg}
  O.~Castillo-Felisola, A.~Melfo, N.~Pantoja and A.~Ramirez,
  %``Localizing gravity on exotic thick 3-branes,''
  Phys.\ Rev.\ D {\bf 70} (2004) 104029
  [arXiv:hep-th/0404083].
  %%CITATION = HEP-TH 0404083;%%


%\cite{Guerrero:2002ki}
\bibitem{Guerrero:2002ki}
  R.~Guerrero, A.~Melfo and N.~Pantoja,
  %``Self-gravitating domain walls and the thin-wall limit,''
  Phys.\ Rev.\ D {\bf 65} (2002) 125010
  [arXiv:gr-qc/0202011].
  %%CITATION = GR-QC 0202011;%%


%\cite{Melfo:2002wd}
\bibitem{Melfo:2002wd}
  A.~Melfo, N.~Pantoja and A.~Skirzewski,
  %``Thick domain wall spacetimes with and without reflection symmetry,''
  Phys.\ Rev.\ D {\bf 67} (2003) 105003
  [arXiv:gr-qc/0211081].
  %%CITATION = GR-QC 0211081;%%


\bibitem{g90}
G.~Goetz,
  J.\ Math. \ Phys. {\bf 31} (1990) 2683.


\bibitem{gassmukh}
  R.~Gass and M.~Mukherjee,
  %``Domain wall spacetimes and particle motion,''
  Phys.\ Rev.\ D {\bf 60} (1999) 065011
  [arXiv:gr-qc/9903012].
  %%CITATION = GR-QC 9903012;%%


 %\cite{Randjbar-Daemi:2000cr}
\bibitem{Randjbar-Daemi:2000cr}
  S.~Randjbar-Daemi and M.~E.~Shaposhnikov,
  %``Fermion zero-modes on brane-worlds,''
  Phys.\ Lett.\ B {\bf 492} (2000) 361
  [arXiv:hep-th/0008079].
  %%CITATION = HEP-TH 0008079;%%

%\cite{Dubovsky:2000am}
\bibitem{Dubovsky:2000am}
  S.~L.~Dubovsky, V.~A.~Rubakov and P.~G.~Tinyakov,
  %``Brane world: Disappearing massive matter,''
  Phys.\ Rev.\ D {\bf 62} (2000) 105011
  [arXiv:hep-th/0006046].
  %%CITATION = HEP-TH 0006046;%%
  %%Cited 47 times in SPIRES-HEP

 %\cite{Ringeval:2001cq}
\bibitem{Ringeval:2001cq}
  C.~Ringeval, P.~Peter and J.~P.~Uzan,
  %``Localization of massive fermions on the brane,''
  Phys.\ Rev.\ D {\bf 65} (2002) 044016
  [arXiv:hep-th/0109194].
  %%CITATION = HEP-TH 0109194;%%


%\cite{Koley:2004at}
\bibitem{Koley:2004at}
  R.~Koley and S.~Kar,
  %``Scalar kinks and fermion localisation in warped spacetimes,''
  Class.\ Quant.\ Grav.\  {\bf 22} (2005) 753
  [arXiv:hep-th/0407158].
  %%CITATION = HEP-TH 0407158;%%


%\cite{Skenderis:1999mm}
\bibitem{Skenderis:1999mm}
  K.~Skenderis and P.~K.~Townsend,
  %``Gravitational stability and renormalization-group flow,''
  Phys.\ Lett.\ B {\bf 468} (1999) 46
  [arXiv:hep-th/9909070].
  %%CITATION = HEP-TH 9909070;%%


\bibitem{Pantoja:2003zr}
  N.~Pantoja and A.~Sanoja,
  %``Symmetries of distributional domain wall geometries,''
  J.\ Math. \ Phys. {\bf 46} (2005) 033509-1-13
  [arXiv:gr-qc/0312032].
  %%CITATION = GR-QC 0312032;%%


%\cite{Kehagias:2002qk}
\bibitem{Kehagias:2002qk}
  A.~Kehagias and K.~Tamvakis,
  %``Graviton localization and Newton law for a dS(4) brane in 5D bulk,''
  Class.\ Quant.\ Grav.\  {\bf 19} (2002) L185
  [arXiv:hep-th/0205009].
  %%CITATION = HEP-TH 0205009;%%


%\cite{Arkani-Hamed:1999dc}
\bibitem{Arkani-Hamed:1999dc}
  N.~Arkani-Hamed and M.~Schmaltz,
  %``Hierarchies without symmetries from extra dimensions,''
  Phys.\ Rev.\ D {\bf 61}, 033005 (2000)
  [arXiv:hep-ph/9903417].
  %%CITATION = HEP-PH 9903417;%%


\bibitem{grmp}
  R. Guerrero, A. Melfo, N. Pantoja and R.O. Rodr\'iguez, work in preparation.


\bibitem{Guerrero:2005aw}
  R.~Guerrero, R.~O.~Rodriguez and R.~Torrealba,
  %``de Sitter and double asymmetric brane worlds,''
  arXiv:hep-th/0510023;
  %%CITATION = HEP-TH 0510023;%%


%\cite{Guerrero:2005xx}
%\bibitem{Guerrero:2005xx}
  R.~Guerrero, R.~Ortiz, R.~O.~Rodriguez and R.~Torrealba,
  %``Irregular BPS domain walls,''
  arXiv:gr-qc/0504080.
  %%CITATION = GR-QC 0504080;%%


\end{thebibliography}
\end{document}